# The 1932 Majorana equation: a forgotten but surprisingly modern particle theory

Luca Nanni

luca.nanni@edu.unife.it

The Standard Model is an up-to-date theory that best summarizes current knowledge in particle physics. Although some problems still remain open, it represents the leading model which all physicists refer to. One of the pillars which underpin the Standard Model is represented by the Lorentz invariance of the equations that form its backbone. These equations made it possible to predict the existence of particles and phenomena that experimental physics had not yet been able to detect. The first hint of formulating a fundamental theory of particles can be found in the 1932 Majorana equation, formulated when electrons and protons were the only known particles. Today we know that parts of the hypotheses set by Majorana were not correct, but his equation hid concepts that are found in the Standard Model. In this study, the Majorana equation is revisited and solved for free particles. The time-like, light-like and space-like solutions, represented by infinite-component wave functions, are discussed. Furthermore, by introducing subsidiary conditions on the mass term, it is possible to quantize both the fermionic and the bosonic towers, obtaining the mass spectrum of the entire family of charged leptons, baryons and mesons.

## 1. Introduction

The formulation of the Majorana relativistic theory of particles with arbitrary spin [1] appeared in 1932 and remained unknown for a long time despite being full of remarkable new ideas. The reasons are attributable to the fact that the paper was published only in Italian; moreover it dealt with a premature problem both with respect to the dominant interests among physicists of the early thirties, and with respect to knowledge of experimental results relating to elementary particles [2-4]. In fact, the work was sent by Majorana to *Il Nuovo Cimento*, before Rome received news of the discovery of the positron announced by Anderson [5]. It is from the forties that Majorana's article began to attract the attention of mathematically oriented theoretical physicists who exploited his ideas, often without ever citing him expressly in their works, and developed theories that helped to build the Standard Model [6-12].

In the Majorana paper, there are the first hints of supersymmetry, of the spin-mass correlation and of spontaneous symmetry breaking: three fundamental conceptual bases of the Standard Model [13]. These hints prove that our conceptual understanding of the fundamental laws of nature were already in Majorana attempts to describe particles with arbitrary spins in a relativistic invariant way. This proves how interesting the Majorana theory still is and why it is worth investigating its formalism. Concerning the first concept, Majorana states that the simplest representation of the Lorentz group is infinite-dimensional. In this representation, the states with integer (bosons) and half-integer (fermions) spins are treated on equal grounds. In other words, the relativistic description of particle states allows bosons and fermions to exist on equal grounds. These two fundamental sets of states are just the first hint of supersymmetry. The second remarkable novelty is the correlation between spin and mass. As we will see later, the mass eigenvalues provided by the Majorana equation are given by $m_J = \chi/(J + 1/2)$, where $J$ is the spin and $\chi$ is a given mass term (whose meaning will be clarified later). The mass decreases with increasing spin, the opposite of what would appear, many decades later, in the study of the strong interactions between baryons and mesons (Chew–Frautschi–Gribov–Regge trajectories) [14]. However, by assigning to $\chi$ a suitable value, the Majorana equation returns the entire mass spectrum of each specific particle. The last remarkable concept hidden in the Majorana equation is that of spontaneous symmetry breaking. In fact, the equation also provides imaginary mass eigenvalues, and today we know that the symmetry-breaking mechanism – which is the only way to introduce the real masses of particles forming the universe we know – could not exist without imaginary masses [15]. There is a further development, which this paper contributed to: the formidable relation between spin and statistics, which leads to the discovery of another invariance law, valid for all quantized relativistic field theories: the CPT theorem. First of all, the Majorana paper shows that the relativistic description of a particle state allows the existence of integer and half-integer spin values. But it was already known that the electron must obey the Pauli exclusion principle and that the electron has half-integer spin. Thus, the problem arises of understanding if the Pauli principle is valid for all half-integer spins. If this were the case, it would be necessary to find which properties characterize these two classes of particles (fermions and bosons). The first of these properties are of a statistical nature, governing groups of identical fermions and groups of identical bosons. We now know that a fundamental distinction exists and that the bases for the statistical laws governing fermions and bosons are the anticommutation relations for fermions and the commutation relations for bosons [16]. Well, the Majorana equation provides the correct algebraic relations, depending on whether the construction of the spin matrices is done for fermions or bosons. We can state that Majorana's work is a precursor of the spin-statistics theorem proved years later by Pauli [17].

Today, almost a century later, what was considered a purely mathematical curiosity in 1932 represents a powerful source of incredibly new ideas such as those mentioned earlier. Since Majorana's equation for particles with arbitrary spins remains unknown to the majority of the physics community, it must be revisited in a modern key, not only for purely historical reasons but also, and above all, because it is a tool that could help solve some of the open questions of particle physics or be useful in other areas of physics, such as quantum optics. This is the goal that this work sets out to achieve. The paper is organised as follows: § 2 is dedicated to the Majorana formalism, based on the theory of Lie groups, which leads to the correct formulation of the commutation relations for fermions and bosons, regardless of the spin value. In § 3 the Majorana equation is solved both by the algebraic approach and by complex analysis. The first method allows us to explicitly obtain the probability of the existence of states with a given spin as a function of the particle energy. The second, on the other hand, brings out the physical meaning of all the possible solutions, including superluminal ones. Finally, in § 4, through the introduction of subsidiary conditions, the mass term, which Majorana never discussed in his work, assumes the value of a universal constant from which the mass spectrum of the known particles can be obtained. In particular, for charged leptons, baryons and mesons, the spectrum is explicitly obtained.

**2. The Majorana Formalism**

Majorana's goal was to find a relativistic equation describing particles with arbitrary spin. The relativistic invariance requires that spatiotemporal coordinates are treated on the same footing, and, using the form of the Dirac equation [18], Majorana writes down its equation as:

$$(i\partial_t - i\alpha^i \partial_i - \beta\chi)\psi = 0, \tag{1}$$

Where the natural units $\hbar = c = 1$ have been used (this notation will be used throughout the article). The index ($i = 1, 2, 3$) is for spatial coordinates, $\alpha^i$ and $\beta$ are numerical matrices to be found, and $\chi$ is a mass term that we leave unspecified for now. Since Majorana considers the negative energy solutions of the Dirac equation to be unmeaningful, Eq. (1) must be formulated so that matrix $\beta$ is positive-definite. Moreover, Majorana does not require the respect of the relativistic energy–momentum relationship. This means that Eq. (1) multiplied times its conjugate does not necessarily have to return the Klein–Gordon equation. Hence, not only are the dimensions of the matrices $\alpha^i$ and $\beta$ different from those of Dirac (at least for spin-1/2 particles), but also commutation relations change. The set of these hypotheses determines the Majorana formalism for the construction of the matrices $\alpha^i$ and $\beta$.

To obtain the explicit form of matrices $\alpha^i$ and $\beta$, Majorana uses the variational principle (see Eq. (2) in the original paper of reference [1].) Using a modern formalism, the same result can be obtained by imposing the Lorentz invariance on the Lagrangian density This Lagrangian reads:

$$\mathfrak{L} = \psi^\dagger(i\partial_t - i\alpha^i \partial_i - \beta\chi)\psi. \tag{2}$$

The invariance of Lagrangian (4) implies that all the terms that form it are relativistically invariant. Therefore, the term $\psi^\dagger \beta \psi$ must also be invariant under the action of the elements of the Lorentz group. This means that it is possible to perform a non-unitary transformation $\psi \to \varphi$ such that $\psi^\dagger \beta \psi \to \varphi^\dagger \varphi$, where the term $\varphi^\dagger \varphi$ must also be relativistically invariant. This transformation is necessary to ensure that the Majorana spinors that being formulated represent the Lorentz group (which, as we will see later, will no longer be a finite representation, as are the Dirac spinors, but rather infinite-dimensional). To avoid mathematical complications, Majorana uses infinitesimal Lorentz transformations in the variables $t, x, y, z$, whose space-space and space-time generators respectively:

$$S_1 = \begin{pmatrix} 0_{2\times 2} & 0_{2\times 2} \\ 0_{2\times 2} & \begin{matrix} 0 & \bar{1} \\ 1 & 0 \end{matrix} \end{pmatrix}, S_2 = \begin{pmatrix} 0_{2\times 2} & \begin{matrix} 0 & 0 \\ 0 & 1 \end{matrix} \\ \begin{matrix} 0 & 0 \\ 0 & \bar{1} \end{matrix} & 0_{2\times 2} \end{pmatrix}, S_3 = \begin{pmatrix} 0_{2\times 2} & \begin{matrix} 0 & 0 \\ \bar{1} & 0 \end{matrix} \\ \begin{matrix} 0 & 1 \\ 0 & 0 \end{matrix} & 0_{2\times 2} \end{pmatrix}, \qquad (3)$$

$$T_1 = \begin{pmatrix} \begin{matrix} 0 & 1 \\ 1 & 0 \end{matrix} & 0_{2\times 2} \\ 0_{2\times 2} & 0_{2\times 2} \end{pmatrix}, T_2 = \begin{pmatrix} 0_{2\times 2} & \begin{matrix} 1 & 0 \\ 0 & 0 \end{matrix} \\ \begin{matrix} 1 & 0 \\ 0 & 0 \end{matrix} & 0_{2\times 2} \end{pmatrix}, T_3 = \begin{pmatrix} 0_{2\times 2} & \begin{matrix} 0 & 1 \\ 0 & 0 \end{matrix} \\ \begin{matrix} 0 & 0 \\ 1 & 0 \end{matrix} & 0_{2\times 2} \end{pmatrix}, \qquad (4)$$

where $0_{2\times 2}$ is the 2x2 zero matrix and $\bar{1} = -1$. The $S_i$ space-space transformations are rotations, while the $T_i$ space-time transformations are boosts. Such transformations have the advantage of simplifying the formulation of the equations, and by their integration it is possible to obtain the corresponding finite transformations. From generators (3) and (4), the following operators are obtained:

$$\begin{cases} \hat{a}_1 = iS_1, \hat{a}_2 = iS_2, \hat{a}_3 = iS_3 \\ \hat{b}_1 = -iT_1, \hat{b}_2 = -iT_2, b_3 = -iT_3 \end{cases}. \qquad (5)$$

The operators $\hat{a}_i$ are Hermitian, whereas operators $\hat{b}_i$ are anti-Hermitian. Notice that while the $\hat{a}_i$ are Hermitian, the boosts $\hat{b}_i$ are anti-hermitian, this being related to the fact that the Lorentz group is non-compact (topologically, the Lorentz group is $\mathbb{R}_3 \times S_3/\mathbb{Z}_2$, the non-compact factor corresponding to boosts and the doubly connected $S_3/\mathbb{Z}_2$ corresponding to rotations). In order to construct a unitary representation $\hat{a}_i$ and $\hat{b}_i$ must be Hermitian, and *vice versa*. To satisfy this requirement and for infinitesimal transformations to be integrable, Majorana introduced appropriate commutation relations. To obtain these, let us consider the following example in $\mathbb{R}^3$. Let $R_1(\alpha)$ and $R_2(\theta)$ be two infinitesimal rotations about the $x$ and $y$ axes respectively, whose exponential representations are:

$$R_1(\alpha) = e^{\alpha S_1} \quad and \quad R_2(\alpha) = e^{\theta S_2}. \tag{6}$$

The consecutive action of the two rotations is given by:

$$R_1(\alpha)R_2(\theta) = e^{(\alpha S_1 + \theta S_2)}. \tag{7}$$

If we develop this equation using Taylor series, we obtained:

$$R_1(\alpha)R_2(\theta) = 1 + (\alpha S_1 + \theta S_2) + \frac{1}{2}(\alpha S_1 + \theta S_2)^2 + \frac{1}{2}[\alpha S_1, \theta S_2] + \cdots, \tag{8}$$

where

$$[\alpha S_1, \theta S_2] = \alpha\vartheta S_1 S_2 - \alpha\vartheta S_2 S_1. \tag{9}$$

The commutator (9) has been introduced in Eq. (8) to obtain the correct form of the square binomial that appears in the integration formulas:

$$(\alpha S_1 + \theta S_2)^2 + [\alpha S_1, \theta S_2] = \alpha^2 S_1^2 + 2\alpha\theta S_1 S_2 + \theta^2 S_2^2. \tag{10}$$

Since in the Taylor series (8) binomials also appear, $(\alpha S_1 + \theta S_2)^n$ with $n > 2$, it is necessary to assure their correct forms in order to introduce further commutators. The procedure is completely analogous to that of the case $n = 2$. However, having used infinitesimal generators, the development (8) can be truncated at the third term, thus avoiding the calculation of the higher-order commutators as well.

The commutator (9) recalls that which is typical of the angular momentum obtained by Schrödinger, Heisenberg and Dirac in their quantum theories [18-20]. Therefore, whatever the combinations between infinitesimal transformations (5), the following commutation relations must hold:

$$[\hat{a}_i, \hat{a}_j] = i\varepsilon_{ijk}\hat{a}_k, [\hat{b}_i, \hat{b}_j] = -i\varepsilon_{ijk}\hat{a}_k, [\hat{a}_i, \hat{b}_j] = i\varepsilon_{ijk}b_k, \tag{11}$$

where $i, j, k = 1,2,3$. The elements of the operators $\hat{a}_i$ and $\hat{b}_i$ in the matrix representation are obtained using the wave functions of the total angular momentum operator, characterized by quantum numbers $j$ and $m$:

$$\begin{cases} \langle j,m|\hat{a}_1 \mp i\hat{a}_2|j,m\pm 1\rangle = [(j\pm m+1)(j\mp m)]^{1/2} \\ \langle j,m|\hat{a}_3|j,m\rangle = m \\ \langle j,m|\hat{b}_1 \mp i\hat{b}_2|j\pm 1,m\pm 1\rangle = -\frac{1}{2}\pm\{[j\pm(m\pm 1)][j+1\pm(m\pm 1)]\}^{1/2}, \\ \langle j,m|\hat{b}_3|j\pm 1,m\rangle = \frac{1}{2}\{[j+m+1][j-m+1]\}^{1/2} \end{cases} \quad (12)$$

where $j$ can take both half-integer and integer values, while $-j \leq m \leq j$. All the non-trivial components of the infinite-dimensional Majorana matrices are obtained by varying $j$ from 0 to $\infty$ through integer steps (see Eq. (9) of reference 1). Therefore, Majorana's approach treats half-integer and integer spin states on equal grounds. This is the first hint of the supersymmetry that was anticipated in the previous section. A closer analysis of relations (12) reveals that operators $\hat{a}_1 \mp i\hat{a}_2$ and $\hat{b}_1 \mp i\hat{b}_2$ are nothing but the creation and annihilation operators, already introduced a few years earlier by Dirac for the harmonic oscillator[18]. In his theory for particles with arbitrary spin, Majorana generalizes this formalism, which is now commonly used in quantum field theory.

We must now find the explicit form of transformation $\psi \to \varphi$. In this regard, let us rewrite Eq. 2 as follows:

$$\mathfrak{L} = \varphi^\dagger(i\gamma^\mu\partial_\mu - \chi)\varphi, \quad (13)$$

where we set $\varphi^\dagger = \beta^{-1}\psi^\dagger$ and $\varphi = \psi$. In this way we obtain:

$$\psi^\dagger\beta\psi = \varphi^\dagger\varphi. \quad (14)$$

To ensure the invariance of the Lagrangian (13), the following commutation relations must hold:

$$[\gamma^0,\hat{a}_1] = 0\,, [\gamma^0,\hat{b}_1] = i\gamma^1\,, [\gamma^1,\hat{a}_1] = 0, [\gamma^1,\hat{b}_2] = i\gamma^3\,, [\gamma^1,\hat{a}_3] = -i\gamma^2\,, [\gamma^1,\hat{b}_1] = i\gamma^0\,,\cdots. \quad (15)$$

It is easy to verify that the explicit form of the matrices $\gamma^\mu$, with $\mu = 0,1,2,3$, is $\gamma^0 = \beta^{-1}$ and $\gamma^i = \beta^{-1}\alpha^i$. The explicit form of the matrices $\gamma^\mu$ can be obtained by proceeding as we did for operators $\hat{a}_i$ and $\hat{b}_j$:

$$\begin{cases} \langle j,m|\gamma^0|j,m\rangle = -i(j+1/2) \\ \langle j,m|\gamma^1 \mp i\gamma^2|j\pm 1, m\pm 1\rangle = \mp\frac{1}{2}\{[j\pm(m\pm 1)][j+1\pm(m\pm 1)]\}^{1/2} \\ \langle j,m|\gamma^3|j\pm 1, m\rangle = \pm\frac{1}{2}\{[j+m+1][j-m+1]\}^{1/2} \end{cases} \quad (16)$$

All other matrix elements different from those in Eq. (16) are trivially zero. We are now able to determine the transformation T such that $\varphi = T\psi$. To obtain more information on the nature of this transformation, let us write:

$$\varphi^\dagger \varphi = (T\psi)^\dagger(T\psi) = \psi^\dagger(T^\dagger T)\psi. \quad (17)$$

The transformation $T$ must be such that $\psi^\dagger \beta \psi = \varphi^\dagger \varphi$, and this holds if $T^\dagger T = \beta$, i.e. $T$ is not unitary. Considering that $\gamma^0 = \beta^{-1}$ and that $\langle j,m|\gamma^0|j,m\rangle = -i(j+1/2)$, we obtain:

$$\langle j,m|T|j,m\rangle = (j+1/2)^{-1/2}. \quad (18)$$

In Eq. (18) the explicit form of matrix $\beta$, given by $\beta = T^2$, is obtained. Moreover, the spin matrices $\alpha^\nu$ are given by $\alpha^\nu = T\gamma^\nu T$. It must be observed that the non-unitary of the transformation T is due to the fact that Majorana does not require compliance with the energy-momentum relation, a necessary condition for $T^2 = \mathbb{1}$. The form of matrix $\beta$ leads to a discrete mass spectrum, determined by the term $\chi$ and the particle spin:

$$m_j = \chi/(j+1/2). \quad (19)$$

In his article Majorana does not comment on this result, considering it trivial. On the other hand, he is mainly interested in the mathematical aspects of his research, aiming to obtain the first infinite-dimensional representation of the Lorentz group, an algebraic formalism unknown to most physicists at that time. Several authors have tried to give a physical explanation to Eq. (19), without ever obtaining relations consistent with the measured values of the masses of the known particles [21-24]. The widespread opinion of those who dealt with the 1932 Majorana equation is that the mass spectrum of Eq. (19) is mostly a side-effect of the theory for particles with arbitrary spin, essentially due to the elimination of the negative energy solutions and the non-respect of the energy–momentum relation. However, the mass term $\chi$ is never made explicit by Majorana, and this leaves open the possibility of investigating other less conventional approaches that could lead to new research perspectives aimed at solving the problem of the masses [25-26]. In this study, we will investigate the hypothesis in which $\chi$ is a sort of universal constant from which, as the spin varies, the entire spectrum of known particle masses is obtained. We will address this topic in § 4.

Majorana's approach leads to the formulations of infinite matrices $\alpha^\mu$ and $\gamma^\mu$, and therefore the solutions of Eq. (1) are wave functions with infinite components. In the reference frame with $\boldsymbol{p} \neq 0$ (where $\boldsymbol{p}$ is the space component of four-momentum $p^\mu$), each component $\psi_{j,m}$ of the wave function $\psi$ is characterized by the quantum numbers $j$ and $m$. For fermions, the form of the wave function is:

$$\psi = (\psi_{1/2,1/2}, \psi_{1/2,-1/2}, \psi_{3/2,3/2}, \psi_{3/2,1/2}, \psi_{3/2,-1/2}, \psi_{3/2,-3/2}, \cdots), \quad (20)$$

while for bosons it is:

$$\psi = (\psi_0, \psi_{1,1}, \psi_{1,0}, \psi_{1,-1}, \psi_{2,2}, \psi_{2,1}, \psi_{2,0}, \psi_{2,-1}, \psi_{2,-2} \cdots). \quad (21)$$

Majorana proves that the probability of existence of states with spin $j$ is proportional to $(u/c)^n$, where $u$ is the particle velocity and $n$ is a positive integer, such that $n = 1$ for $j = 1/2$ or $0$, $n = 2$ for $j = 3/2$ or $1$, and so on. As $u \ll c$, the probability that states with high $j$ exist decreases progressively. Therefore, it is reasonable to assume that the high spin states are excited states of other particles accessible only in ultrarelativistic regimes.

The Majorana equation also provides space-like solutions, and this too is a direct consequence of not respecting the energy-momentum relation. Today we know that the only way to introduce real masses, without destroying the theoretical description of nature, is by the mechanism of spontaneous symmetry breaking, and such a mechanism could not exist without the imaginary masses. Almost a century ago Majorana had formulated such a sophisticated theory that, had it been published in an international journal, would probably have accelerated the construction of modern particle theory.

Before going to the next section, it is appropriate to study how the Majorana equation behaves under the action of CPT symmetry, considered to be the only exact discrete symmetry in nature. The wave operator of Eq. (1) is CPT invariant if the transformations $p_\mu \to -p_\mu$ and $\Gamma^\mu \to -\Gamma^\mu$ hold (the operator $\Gamma^\mu$ is the vector whose components are the $\gamma^\nu$ matrices). The transformation $\Gamma^\mu \to -\Gamma^\mu$ is obtained by applying a rotation $\mathfrak{R}_z(\pi)$ of $\pi$ along the $z$ axis followed by a boost $\mathfrak{B}_z(i\pi)$ of imaginary parameter $i\pi$ along the same direction. For Dirac 1/2-spin particles, the product of these two consecutive transformations gives the matrix $\gamma^5$, but in the case of Majorana particles, whatever the spin, this relation is not more valid since the eigenvalues of $\gamma^0$ are all positive. In the framework of infinite unitary representation of the Lorentz group, the parameter $\xi = i\pi$ is a pole for the boost $\mathfrak{B}_z$. This is the mathematical reason why the transformation $\Gamma^\mu \to -\Gamma^\mu$ does not exist for Eq. (1). This also compromises the possibility of deriving the spin-statistics theorem from the Majorana equation. However, it cannot be a priori ruled out that Majorana's theory can be modified to make it invariant under the action of CPT symmetry. This is one of the points worth working on, and in any case the goal is to make Majorana's theory complementary to the Standard Model, not to replace it.

**3. Solving the Majorana Equation for Free Particles**

Solving the Majorana equation is a challenging and fascinating task. This equation corresponds to an infinite number of linear equations in an infinite number of variables, a problem that has been a classic of contemporary mathematics. An exhaustive discussion of this problem can be found in reference [27]. For our purposes, let us rewrite the Majorana equation as:

$$(i\gamma^0 \partial_t - i\gamma^i \partial_i - \chi)\varphi = 0, \tag{22}$$

where, as usual, $i = 1,2,3$. The wave function $\psi$ is the infinite-component spinor of the Majorana equation to be found. One of the constraints which this equation must satisfy is the normalization condition $\langle \varphi | \varphi \rangle = 1$, which is the usual dot product between eigenvectors in the Dirac formalism, to have $\varphi$ the meaning of probability amplitude. This condition is one of the requisites for making Eq. (22) admit a non-trivial solution [27]. The explicit forms of matrices $\gamma^0$ and $\gamma^i$ were obtained in the previous section. In particular, matrix $\gamma^0$ has all zero elements except those on the main diagonal, which are real and positive. On the other hand, matrix $\gamma^3$, unlike $\gamma^1$ and $\gamma^2$, is a block off-diagonal matrix. Matrices $\gamma^1$ and $\gamma^2$ have non-trivial elements on the secondary diagonals adjacent to the main one. Let us rewrite Eq. (22) as:

$$\left(i\frac{1}{\chi}\gamma^0\partial_t - i\frac{1}{\chi}\gamma^i\partial_i - \mathbb{1}\right)\varphi = 0 \quad \Rightarrow \quad A\varphi - \mathbb{1}\varphi = 0, \tag{23}$$

where $A$ is the matrix given by the sum of the terms in $\partial_t$ and $\partial_\nu$. Eq. (23) can also be rewritten as $\varphi^{-1}A\varphi = \mathbb{1}$, which is the typical similitude relation of linear algebra. It is proved that Eq. (23) can be solved if $A$ is a lower half-matrix or a block matrix which develops on the main diagonal [28]. In the case under study, whatever the integer or half-integer value of the spin is, $A$ is always a block matrix constructed along the main diagonal. This property ensures that Eq. (23) admits a non-trivial solution. Its Hamiltonian is given by $\mathbb{H} = \sum_i^3 \gamma^i \hat{p}_i + \chi(\gamma^0)^{-1}$, where $\hat{p}_i$ are the impulse operators and $(\gamma^i, \gamma^0)$ are the spin matrices constructed as explained above. This Hamiltonian is an infinite sequence.

Following this approach, and considering that matrix $\varphi$ has non-trivial $\varphi_\mu^{(j)}$ elements only on the main diagonal, we obtain the following proportions:

$$\begin{cases} \varphi_r^{(j)} \sim \left(\pm\frac{p_3}{E+\chi}\right)^{j+1/2} , & \varphi_s^{(j)} \sim \left[\frac{(p_1 \pm ip_2)}{E+\chi}\right]^{j+1/2} \quad for\ j = n/2 \\ \varphi_r^{(j)} \sim \left(\pm\frac{p_3}{E+\chi}\right)^{j+1} , & \varphi_s^{(j)} \sim \left[\frac{(p_1 \pm ip_2)}{E+\chi}\right]^{j+1} \quad for\ j = n \end{cases}, \tag{24}$$

where $r$ and $s$ are indexes related by $s = r + 1$, while $p_\nu$ are the spatial components of four-vector $p_\mu$. For time-like solutions, the term $p_\nu/E + \chi$ is always lower than one. This means that as the spin increases, the components $\varphi_r^{(j)}$ and $\varphi_s^{(j)}$ become progressively smaller, tending to zero as $j \to \infty$. This convergence is the second requirement needed to ensure that Eq. (23) is solvable [27-28]. The spinor components that contribute the most to the calculation of the average value $\langle\varphi|\hat{O}|\varphi\rangle$ of a generic observable O, where $\hat{O}$ is the Hermitian operator corresponding to this variable, are the ones with small $j$. The relevance of high spin components increases as the particle velocity increases. It has been proved that the probability $P(n)$ of the existence of a state with half-integer spin $j$ is [29]:

$$P(n) = [(u)^n - (u)^{n+1}] \quad for\ j = n/2, \qquad (25)$$

where $n$ is the order of the state under consideration (in his original work, Majorana maintains the physical meaning of $\varphi$ as a probabilistic wave function, in line with the previous theories of Dirac and Schrodinger). For luxons, all spin states become equally probable, while for space-like particles the probability of existence increases as $j$ increases, diverging as $j \rightarrow \infty$. This compromises our ability to solve Eq. (23) for these specific cases, at least by using the algebraic approach. However, this difficulty can be overcome by addressing the solution of the Majorana equation by means of complex analysis.

The complex analysis approach to solve the Majorana equation was proposed by Barut [30]. This approach, besides being simpler and more powerful than the algebraic one, has the advantage of better highlighting the physical meaning of the solutions. The method of complex analysis is based on the fact that once a solution of the equation is obtained for a particular configuration, it is possible to obtain all the others by means of Lorentz transformations. Let us suppose $\varphi$ is a known solution of Eq. (22) for a given configuration, and we want to obtain the solution $\varphi'$ for a configuration with different spin. Therefore, a transformation $U$ must be found, depending on the Lorentz transformation $\Lambda$, such that:

$$\varphi' = U(\Lambda)\varphi \quad \Rightarrow \quad \varphi = U^{-1}(\Lambda)\varphi'. \qquad (26)$$

If we substitute Eq. (26) for Eq. (22), we obtain:

$$\left(i\gamma^0 \partial_t - i\gamma^i \partial_i - \chi\right) U^{-1}(\Lambda) \varphi' = 0. \qquad (27)$$

By multiplying on the left side with $U(\Lambda)$, we obtain:

$$\left(iU\gamma^0 U^{-1} \partial_t - iU\gamma^i U^{-1} \partial_i - \chi\right) \varphi' = 0, \qquad (28)$$

which has the same form of Eq. (22).

Let us start analysing the case of time-like solutions. The simplest configuration is that of the centre-of-mass reference frame, for which the Majorana equation becomes:

$$(i\gamma^0 \partial_t - \chi)\varphi = 0. \tag{29}$$

Considering that the eigenvalues of matrix $\gamma^0$ are given by $(j + 1/2)$, the mass spectra associated with $\varphi$ is $\chi/(j + 1/2)$. Let us introduce representation of the Lorentz group through complex analysis [31-32], where the Lie rotations are represented by complex differential operators:

$$z = x + iy, \quad \bar{z} = x - iy, \quad \partial = \frac{\partial}{\partial z}, \quad \bar{\partial} = \frac{\partial}{\partial \bar{z}}. \tag{30}$$

Using these transformations, the four-vector $(\gamma^0, \gamma^1, \gamma^2, \gamma^3)$ can be written as:

$$(\gamma^0, \gamma^1, \gamma^2, \gamma^3) = \left[\frac{1}{2}(z\bar{z} - \partial\bar{\partial}), -\frac{1}{4}(z\bar{\partial} + z\partial), \frac{1}{4}i(z\bar{\partial} + \bar{z}\partial), -\frac{1}{2}(z\bar{z} + \partial\bar{\partial})\right]. \tag{31}$$

The transformations of Eq. (30) allow us to pass from the four-dimensional space to the two-dimensional space. In the latter the operators appear as:

$$z = \frac{x + iy}{\lambda_0}, \quad z\bar{z} = \frac{x^2 + y^2}{\lambda_0^2}, \quad \partial\bar{\partial} = \frac{1}{4}\lambda_0\left(\frac{\partial^2}{\partial x^2} + \frac{\partial^2}{\partial y^2}\right), \tag{32}$$

where $\lambda_0 = (x^2 + y^2)^{1/2}$ is a unit length introduced to obtain dimensionless quantities. If we substitute the operators of Eq. (32) in Eq. (31), we obtain the corresponding functional operator of matrix $\gamma^0$:

$$\gamma^0 = \frac{1}{2}(z\bar{z} - \partial\bar{\partial}) = \frac{1}{2}\left[\frac{x^2 + y^2}{\lambda_0^2} - \frac{1}{2}\lambda_0\left(\frac{\partial^2}{\partial x^2} + \frac{\partial^2}{\partial y^2}\right)\right]. \tag{33}$$

If we substitute Eq. (33) for Eq. (29), we obtain:

$$\left[-\frac{\hbar^2}{2\mu}\left(\frac{\partial^2}{\partial x^2} + \frac{\partial^2}{\partial y^2}\right) + \frac{1}{2}\mu\omega^2\lambda^2\right]\varphi = E\varphi, \tag{34}$$

where:

$$\omega = \frac{2}{\mu\lambda_0^2}, \quad E = \frac{2}{\mu\lambda_0^2}. \tag{35}$$

In Eq. (34), $\mu$ is the reduced mass given by $\mu = \chi/2(j + 1/2)$. Unexpectedly, the Majorana equation of a time-like particle in the centre-of-mass reference frame is equivalent to the Schrödinger equation of a two-dimensional oscillator in an attractive harmonic potential. The values of the masses forming the oscillator are $\chi/(j + 1/2)$ and $\chi/(j + 3/2)$ respectively. Therefore, the solution of Eq. (29) is an infinite set of harmonic oscillators with quantized reduced mass $\mu(j)$. It follows that energy $E$ and frequency $\omega$ are also quantized. In particular, the frequency and energy progressively increase as the spin. This result restores the order in the theory under study: although they have a very small mass, high spin particles are energetically unstable (their energy increases progressively as $j$ increases, as prescribed by Eq. (35)), and this is what is observed in nature. This result complies with the one obtained by Afkhami-Jeddi et al., who proved that there are universal bounds on theories with higher spin massive particles [33]. Hidden between the lines of a cryptic article, this proof had already been shown by Majorana nearly a century ago. The explicit form of the solution of Eq. (34) is:

$$\varphi_j = \mathfrak{p}(x,y) \exp[-\chi\omega(x^2 + y^2)/4(j+1)], \tag{36}$$

where $\mathfrak{p}(x,y)$ is a polynomial in the $x$ and $y$ coordinates. Owing to the particular configuration chosen, in which all the spatial components of the impulse are zero, the components of the spinor depend only on spin $j$ and not on the quantum number $m$.

Let us now consider the massive light-like solutions. In this case the simplest configuration must contain at least a non-zero component of impulse. Choosing the one along the $z$ axis, Eq. (22) becomes:

$$(i\gamma^0 \partial_t \pm i\gamma^3 \partial_z - \chi)\varphi = 0. \tag{37}$$

Since the particle velocity is $u = 1$, it is convenient to use the parametrization $(i\partial_t)\varphi = (i\partial_z)\varphi = \omega$. Moreover, since the matrices $\gamma^0$ and $\gamma^3$ are diagonal, the energy spectrum of Eq. (37) is:

$$\omega = \frac{\chi}{\left(s + \frac{1}{2}\right) \pm m}, \tag{38}$$

where $m$ are the eigenvalues of $\gamma^3$. For this configuration, the spinor components will depend not only on $j$ but also on the quantum number $m$. If we substitute Eq. (31), Eq. (32) and Eq. (38) for Eq. (37), we obtain:

$$\begin{cases} \left[-\frac{\hbar^2}{2\mu}\left(\frac{\partial^2}{\partial x^2} + \frac{\partial^2}{\partial y^2}\right) - \frac{2\chi}{\mu\omega\lambda_0^2}\right]\varphi = 0, \\ \left(\omega\frac{x^2 + y^2}{\lambda_0^2} - \chi\right)\varphi = 0, \end{cases} \tag{39}$$

where the first and second equations correspond to the + and − signs respectively in front of the $\gamma^3$ matrix in Eq. (37). The explicit form of $\omega$ is the same as that given in Eq. (35). The first of Eq. (39) is the Schrödinger equation for a particle in a constant potential and with zero total energy. However, light-like particles cannot have zero energy, and the obtained result can be interpreted by assuming that the first of Eq. (39) describes the motion of a composite system with zero total energy. The second of Eq. (39) corresponds to the Cartesian equation of a circle of radius $\lambda_0^2 \chi/\omega$, which implies that the motion of the harmonic oscillator associated with each spin is constrained to lie on a circle.

Finally, let us consider space-like solutions. The simplest configuration is that in which the particle has zero energy, i.e. the reference frame with infinite velocity. For simplicity, we assume that only the $z$ component of space-like momentum is different from zero. Eq. (22) then becomes:

$$(i\gamma^3 \partial_z - \chi)\varphi = 0 \quad \Rightarrow \quad m_j^2 = -(\chi/m)^2, \tag{40}$$

where $\chi$ is an imaginary rest mass, $m$ are the eigenvalues of the $\gamma^3$ matrix and $m_j^2$ is the mass spectrum. As usual, if we substitute Eq. (31) and Eq. (32) for Eq. (40), we obtain:

$$\left[-\frac{\hbar^2}{2\mu}\left(\frac{\partial^2}{\partial x^2} + \frac{\partial^2}{\partial y^2}\right) - \frac{1}{2}\mu\omega^2\lambda^2 - E\right]\varphi = 0. \tag{41}$$

Eq. (41) is completely analogous to Eq. (34) except for the repulsive parabolic potential. Therefore, in the framework of Majorana theory, the space-like solutions are energetically unstable but still possible. In particular, this instability decreases as $j$ increases. Subluminal and superluminal particles therefore have a mirror behaviour.

## 4. The Problem of Particle Masses

As is well known, neither classical nor quantum theories have explained the nature and the numerical values of elementary particle masses [34-35]. What is known is that the mass spectrum of elementary particles is discrete, and, according to modern theories, all elementary particles are the excited states of a small set of fundamental particles, which represent the lowest level of this spectrum. In particular, it seems that the discreteness of the mass spectrum is similar to the spectrum of excitation energies of atoms. In the framework of the Standard Model, the Higgs boson is the particle that gives all other fundamental particles mass. But, despite the work of thousands of researchers, it is not clear exactly how it works and why some particles are more massive than others. Therefore, the problem of particle masses remains open. In this scenario, we want to investigate whether the Majorana equation can contribute to filling this knowledge gap, at least as regards the numerical values of the particle masses. The starting point for this attempt is represented by the mass term $\chi$, which, as previously mentioned, is never discussed by Majorana. The hypothesis is that this term may be associated with a kind of universal natural frequency of a harmonic oscillator, whose resonances correspond to all possible values of the particle rest masses. Therefore, the model we will refer to is the one described by Eq. (34). To obtain this universal constant, it is necessary to use empirical mass formulas, and fortunately these are available with good degrees of accuracy for charged leptons, baryons and mesons [36-39]. Therefore, we are introducing a subsidiary condition (or constraint) to the Majorana theory, which is the only way to make sense of the mass spectrum.

Let us start with the lepton sector, using the Nambu-Barut empirical formula [36]:

$$m_l = m_e \left(1 + \frac{3}{2}\alpha^{-1}\sum_{n=0}^{N} n^4\right) \quad N\epsilon\mathbb{N}, \tag{42}$$

where $m_l$ is the lepton mass spectrum, $m_e$ is the electron rest mass and $\alpha$ is the electromagnetic constant. Eq. (42) is based on the idea that to obtain the mass of heavy lepton, a magnetic quantized energy $\frac{3}{2}\alpha^{-1}\sum_{n=0}^{N} n^4$ must be added to the electron rest energy. For $n = 0, n = 1$ and $n = 2$, Eq. (42) returns the electron mass, the muon mass and the tau mass respectively. Using the second of Eq. (35) and considering that in the centre-of-mass frame the rest energy is $m_j c^2$, the Majorana mass spectrum becomes:

$$m_j = 2\frac{1}{\mu \lambda_0^2}. \tag{43}$$

If we recall that $\mu = \chi/2(j + 1/2)$, Eq. (43) can be rewritten as:

$$m_j = 4\frac{\hbar^2}{\chi \lambda_0^2}(j + 1/2). \tag{44}$$

Eq. (44) relates to Eq. (19) through the equality $\chi/\left(j + \frac{1}{2}\right) = 2/\lambda_0$, obtained by comparing the two equations. This strengthens the hypothesis introduced at the beginning of this section, namely that the mass of the particles is determined by one of the resonant frequencies (and therefore by the single specific wavelengths) of a universal harmonic oscillator. Let us suppose that the masses of the electron, muon and tau correspond to $j = 1/2, j = 3/2$ and $j = 5/2$ respectively. We immediately clarify that these values are not the lepton's spin whose mass spectrum we want to calculate. Their spin is always equal to $1/2$. The values $j = 1/2, 3/2, 5/2$ are obtained from the formula $j = (1/2 + n)$, with $n$ natural number, and represent a new quantum number necessary for quantizing the lepton masses. Specifically, $j = 1/2$ is the quantum number associated with the fundamental frequency of the universal oscillator (mentioned at the beginning of this section), $j = 3/2$ is the value of the quantum number associated with the first order of resonance, $j=5/2$ is the value of the quantum number associated with the second order of resonance, and so on. Following this approach, we can state that the spinor with infinite Majorana components for Fermions is a mathematical object that contains the physical information of the *mother* particle (associated to the universal oscillator) to which all the lepton masses are connected. In other words, the Majorana spinor can be rethought not as the spinor of particles with arbitrary spin but as the spinor of all resonant frequencies of a universal harmonic oscillator. Let us now return to the detail of the calculation we had undertaken. In order for $\chi$ to be constant, $\lambda_0^2$ must depend on $j$, and the product $\lambda_j\sqrt{m_j}$ must be constant:

$$\lambda_{1/2}\sqrt{m_e} = \lambda_{3/2}\sqrt{m_\mu} = \lambda_{5/2}\sqrt{m_\tau}. \tag{45}$$

By using Eq. (42) and Eq. (45), we obtain:

$$\frac{\lambda_{1/2}}{\lambda_{3/2}} = \sqrt{1 + \frac{3}{2}\alpha^{-1}} \quad ; \quad \frac{\lambda_{1/2}}{\lambda_{5/2}} = \sqrt{1 + 24\alpha^{-1}}. \tag{46}$$

If we introduce the Compton wavelength of electron, muon and tau, denoted by $\lambda_{C_e}$, $\lambda_{C_\mu}$ and $\lambda_{C_\tau}$ respectively, we obtain:

$$\lambda_{1/2} = \lambda_{C_e} \quad ; \quad \lambda_{3/2} = 2\lambda_{C_e} \quad ; \quad \lambda_{5/2} = 6\lambda_{C_e}. \tag{47}$$

By using Eq. (46) and Eq. (47), the explicit forms of $\lambda_{3/2}$ and $\lambda_{5/2}$ are obtained:

$$\lambda^2{}_{3/2} = 2\lambda^2{}_{C_e} / \left(1 + \frac{3}{2}\alpha^{-1}\right) \quad ; \quad \lambda^2{}_{5/2} = 3\lambda^2{}_{C_e} / (1 + 24\alpha^{-1}), \tag{48}$$

and the mass spectrum of the three charged leptons becomes:

$$m_e = \frac{4}{\chi\lambda^2{}_{C_e}} \quad ; \quad m_\mu = \frac{4}{\chi\lambda^2{}_{C_e}}\left(1 + \frac{3}{2}\alpha^{-1}\right) \quad ; \quad m_\mu = \frac{4}{\chi\lambda^2{}_{C_e}}(1 + 24\alpha^{-1}). \tag{49}$$

From the first of Eq. (44), the mass $\chi$ can be obtained, and its calculated value is 2.021 MeV/$c^2$, which is very close to the mass value of the up quark of $m_u = 2.3^{+0.7}_{-0.5}\ MeV/c^2$. This value is the universal constant for the lepton sector, limited to charged particles.

Let us consider the baryon and meson sector. To this purpose, an empirical formula has been proposed by Sidharth which reads [39]:

$$m = 137p\left(q + \frac{1}{2}\right), \tag{50}$$

where 137 MeV/$c^2$ is the pion mass, and $p$ and $q$ are positive integer numbers. The formula (50) reproduces the entire mass spectrum of baryons and mesons with errors never exceeding 3%. Sidharth suggested that the numbers $p$ and $q$ could be quantum numbers of harmonic oscillators, without giving proof. Following this idea, and considering that the time-like solutions of the Majorana equation can be interpreted in terms of the energy spectrum of harmonic oscillators with varying reduced mass, we can think of correlating the formula of Eq. (50) with the expressions of Eq. (43). In the centre-of-mass frame, the Majorana oscillator energy is:

$$E = m_j = \frac{2}{\mu \lambda_0^2} \Rightarrow m_j = \frac{2}{\mu \lambda_0^2}, \tag{51}$$

and if we substitute $\mu = \chi/2(j + 1/2)$ in Eq. (51), we obtain the mass spectrum:

$$m_j = \frac{2(j+1)}{\chi \lambda_0^2}. \tag{52}$$

By equating this expression and the formula of Eq. (50), we obtain the algebraic relation between the spin $j$ and the positive integer numbers $p$ and $q$, besides providing physical meaning to the rest mass $\chi$:

$$j = pq + \frac{1}{2}p - 1, \quad \chi = \frac{4}{137\lambda_0^2}. \tag{53}$$

If we assume that the order of magnitude of $\lambda_0$ is no greater than $10^{-15}$ m (i.e. the average dimension of a nucleus), then $\chi$ is of the order of 1 GeV. From the first of Eq. (53), we see that half odd integer spin can be obtained only if $p$ is an odd positive integer. By combining all the possible values of $p$ and $q$, all spin sequences are obtained. It should be noted that the number 137 appears both in Eq. (49) and in Eq. (53), in the first as $\alpha^{-1}$ (of an electromagnetic nature) and in the second as pion mass (of a nuclear nature).

The interpretation we have given to the mass spectrum could be a little speculative and forced, but the logical connection between the empirical formulas and the fact that the Majorana equation can be reformulated as those of quantum harmonic oscillators is evident. Therefore, our speculation still has a robust scientific basis.

## 5. Concluding Discussion

The problem of infinite-dimensional representations of the Lorentz group was discussed in 1938 by Wigner [40], in 1945 by Dirac [41] and again in 1948 by Wigner [42]. Dirac seemed to have been unaware of Majorana's paper, while Wigner, in his book on the Poincaré group, mentioned it almost incidentally. The many studies carried out subsequently by other authors on the infinite-dimensional representations of the Lorentz group did not refer to Majorana's paper. It is not even mentioned by those authors [43] who in recent years have reconstructed, discussed and generalized Majorana theory without realizing it. An exception is Barut and Kleinert's paper [44] which makes specific reference to Majorana results. In this scenario we could say that the 1932 Majorana theory was the object of unconscious plagiarism by many authoritative physicists of the twentieth century.

Apart from the mathematical and historical interest, this work demonstrates Majorana's sensitivity to fundamental problems that he addressed independently, well ahead of most contemporary physicists. With this study, we want to highlight that the Majorana theory for particles with arbitrary angular momentum was not only ahead of its time, but is still so contemporary as to be useful for addressing the study of unsolved problems in particle physics, such as that relating to the mass spectrum, or to investigate other branches of physics with less conventional tools than those commonly used. Reinterpreting Majorana's equation of 1932 and solving it for those physical systems that currently represent open problems could be a way out of the impasse slowing down some sectors of research in particle physics. In the 2006 lecture held at the Majorana Centenary Celebrations symposium, Nambu commented positively on the idea of revisiting the 1932 Majorana equation in view of its application to controversial problems of particle physics. This study hopes to provide an impulse for a physics capable of exploring and approaching problems in an unconventional way, going beyond current theories.

**Conflict of Interest**

The authors declare that they have no conflicts of interest.